\documentclass[useAMS,usenatbib]{mn2e}

\usepackage{epsfig}
\usepackage{longtable}
\usepackage{times}

\usepackage{amsmath}

\def \inte {$INTEGRAL$}

\def \sw {$Swift$}
\def \xte {$RXTE$}

\def \suz {$Suzaku$}
\def \src {IGR\,J16479--4514}

\def \ferg {erg cm$^{-2}$ s$^{-1}$}
\def \hcm {\hbox {\ifmmode $ atom cm$^{-2}\else atom cm$^{-2}$\fi}}

\def \chisq {$\chi ^{2}$}

\def \ATel {Astron.\ Tel.}
\def \apj {ApJ}
\def \apjl {ApJL}
\def \apjs {ApJS}
\def \aap {A\&A}

\def \pasj {PASJ}

\def \mnras {MNRAS}

\def \ssr {Space Science Reviews}

\newcommand{\be}{\begin{equation}}

\newcommand{\ee}{\end{equation}}

\newcommand{\msun}{{~M_{\odot}}}

\title[Suzaku observes one orbit of \src]{A $Suzaku$ X--ray observation of one orbit of the
supergiant fast X--ray transient \src}
\author[L. Sidoli, et al.]{L.\ Sidoli,$^{1}$\thanks{E-mail: sidoli@iasf-milano.inaf.it} P.\ Esposito,$^{1}$  V.\ Sguera,$^{2}$  A.\ Bodaghee,$^{3}$  
J.A.\ Tomsick,$^{3}$ K.\ Pottschmidt,$^{4, 5}$
\newauthor J.\ Rodriguez,$^{6}$   P.\ Romano,$^{7}$ and J.\ Wilms,$^{8}$  \\
$^{1}$INAF, Istituto di Astrofisica Spaziale e Fisica Cosmica,
	Via E.\ Bassini 15,   I-20133 Milano,  Italy \\ 
$^{2}$INAF, Istituto di Astrofisica Spaziale e Fisica Cosmica,
       Via Gobetti 101, I-40129 Bologna, Italy \\ 
$^{3}$Space Sciences Laboratory, 7 Gauss Way, University of California, Berkeley, CA 94720, USA \\
$^{4}$CRESST and NASA Goddard Space Flight Center, Astrophysics Science Division, Code 661, Greenbelt, MD 20771, USA \\ 
$^{5}$Center for Space Science and Technology, University of Maryland Baltimore County, 1000 Hilltop Circle, Baltimore, MD 21250, USA \\
$^{6}$Laboratoire AIM, CEA/IRFU - Universit\'{e} Paris Diderot - CNRS/INSU, CEA DSM/IRFU/SAp, Centre de Saclay, F-91191 Gif-sur-Yvette, France \\
$^{7}$INAF, Istituto di Astrofisica Spaziale e Fisica Cosmica, 
Via U. La Malfa 153, I-90146 Palermo, Italy \\
$^{8}$Dr. Karl Remeis-Sternwarte and Erlangen Centre for Astroparticle Physics, 
Freidrich-Alexander-Universit\"{a}t Erlangen-N\"{u}rnberg,
Sternwartstra$\beta$e 7, \\ 
96049 Bamberg, Germany 
}

\begin{document}

\date{Accepted 2012 December 4.  Received 2012 November 20; in original form 2012 October 10}

\pagerange{\pageref{firstpage}--\pageref{lastpage}} \pubyear{2012}

\maketitle

\label{firstpage}

\begin{abstract}

We report on a 250~ks long X--ray observation
of the supergiant fast X--ray transient (SFXT) \src\
performed with \suz\  in 2012 February.
During this observation, about 80\% of the short orbital period (P$_{orb}$$\sim$3.32 days) 
was covered as continuously as possible for the first time. 
The source light curve displays variability of more than two orders of
magnitude, starting with a very low emission state (10$^{-13}$~\ferg; 1--10 keV) 
lasting the first 46 ks, consistent with being due to the X--ray eclipse by the supergiant companion.
The transition to the uneclipsed X--ray emission is energy dependent. 
Outside the eclipse, the source spends most of the time at a level of 6-7$\times10^{-12}$~\ferg\
punctuated by two structured faint flares
with a duration of about 10 and 15 ks, respectively, reaching a peak flux
of 3-4$\times10^{-11}$~\ferg, separated by about 0.2 in orbital phase.
Remarkably, the first faint flare occurs at a similar orbital phase of the
bright flares previously observed in the system.
This indicates the presence of a phase-locked large scale structure in the
supergiant wind, driving a higher accretion rate onto the compact object.
The average X--ray spectrum is hard and highly absorbed, with a column density, N$_{\rm H}$, of 
10$^{23}$~cm$^{-2}$, clearly in excess of the interstellar absorption. 
There is no evidence for variability of the absorbing column density, except that during the eclipse,
where a less absorbed X--ray spectrum is observed.
A narrow Fe~K${_\alpha}$ emission line at 6.4 keV is viewed along the whole orbit, 
with an intensity which correlates with the continuum emission above 7 keV.
The scattered component visible during the X--ray eclipse 
allowed us to directly probe the wind density at the orbital separation, 
resulting in $\rho$$_{\rm w}$=7$\times$10$^{-14}$~g~cm$^{-3}$. 
Assuming a spherical geometry for the supergiant wind, the derived wind density translates into a
ratio $\dot{M}_{w}/ v_{\infty}=7\times$10$^{-17}$~M$_{\odot}$/km which,  
assuming  terminal velocities in a large range 500--3000~km~s$^{-1}$, implies an accretion luminosity 
two orders of magnitude
higher than that observed. As a consequence,
a mechanism should be at work reducing the mass accretion rate. Different possibilities are discussed.

\end{abstract}


	\section{Introduction\label{intro}}

\src\ is a hard X--ray transient discovered  by \inte\ on 2003, August 8--10 \citep{Molkov2003:16479-4514} 
in the energy range 18--50 keV. 
Several X--ray flares were caught by \inte/IBIS, displaying
variable durations (from 0.5 to 50~hr) and peak fluxes (from 20 to about 600 mCrab, 20--60 keV;
\citealt{Sguera2005, Sguera2006, Sguera2008:16479},
\citealt{Walter2007},  \citealt{Ducci2010}).
Recurrent outbursts were also observed by the \sw\  satellite
(\citealt{Kennea2005:16479-4514}, \citealt{Markwardt2006:16479-4514},
\citealt{Romano2008:sfxts_paperII}, 
\citealt{LaParola2009:atel1929}, \citealt{Bozzo2009}).

The proposed  counterpart (2MASS~J16480656-4512068, \citealt{Walter2006})
was confirmed by an accurate localization obtained with $Chandra$ (\citealt{Ratti2010})
and classified as a late O-type supergiant with a 
spectral type O8.5I located at a distance of 4.9~kpc (\citealt{Rahoui2008}, \citealt{Chaty2008}) 
or a O9.5~Iab star located at 2.8$_{-1.7} ^{+4.9}$~kpc \citep{Nespoli2008}.
The optical identification confirmed the initial classification of \src\
as a member of the new sub-class of high mass X--ray binaries, 
the supergiant fast X--ray transients (SFXTs),
at first suggested only based on the 
short duration of its hard X--ray activity \citep{Sguera2006}.

\src\ is an eclipsing SFXT \citep{Bozzo2008:eclipse16479} and the one with the 
narrowest orbit, showing an orbital period
of 3.32~days \citep{Jain2009:16479}, later refined to 3.3193$\pm{0.0005}$\,days \citep{Romano2009:sfxts_paperV}.

X--ray long-term monitoring outside outbursts revealed that the source spends most of its
time at a reduced level of X--ray emission (10$^{33}$-10$^{34}$~erg~s$^{-1}$), 
from 2 to 3 orders of magnitude less
than the  flare peaks, 
while for the remaining $\sim$19\% of the time, it is undetected,
compatibly with being in eclipse (\citealt{Sidoli2008:sfxts_paperI},
\citealt{Romano2009:sfxts_paperV}).

The nature of the compact object is unclear (as in about a half of the members of the SFXTs class; for a recent
review see \citealt{Sidoli2011texas})
but there is indirect evidence based on the spectral similarity with
accreting pulsars suggesting that the X--ray source is a neutron star (NS).

Here we report on the first 
X--ray observation which continuously (except that during the interruptions
because of the satellite orbit) covers most of an orbital cycle of \src.
Our main goal is to investigate the  variability of the X--ray properties
along a single orbital cycle.

 	 \section{Observations and Data Reduction\label{dataredu}}


\src\ was observed  by \emph{Suzaku} \citep{Mitsuda2007} between 2012 February 23 and 26. 
The observation (Obs ID 406078010) was performed with the source located at 
the X--ray Imaging Spectrometer (XIS; \citealt{Koyama2007}) nominal position. 
The XIS consists of four telescopes with a spatial resolution of about 2 arcmin coupled to 
four CCD cameras operating in the 0.2--12 keV energy range and with $18'\times18'$ field of view 
($1024\times1024$ pixels; $1.05$ arcsec pixel$^{-1}$; Koyama et al. 2007).  
At the time of our observation only the two front illuminated (FI) XIS0 and XIS3 
(effective area: 330 cm$^2$ at 1.5 keV), and the back-illuminated (BI) XIS1 (effective area: 370 cm$^2$ at 1.5 keV) 
were operating. 
The three detectors were operated in Normal Mode with no window or burst option 
(all the pixels on the CCD are read out every 8 seconds). 
The other instrument aboard \emph{Suzaku} is the Hard X-ray Detector (HXD; \citealt{Takahashi2007}), 
which consists of PIN silicon diodes (HXD-PIN; 10--70 keV) and Gd$_2$SiO$_5$Ce (GSO; 50--600 keV) scintillators. 
The HXD was operated in the standard mode, with a time resolution of 61 $\mu$s. 
The HXD is a collimated instrument and the PIN in particular has a $67.6'\times67.6'$ 
field of view (FWHM: $34'\times34'$).

Data reduction and analysis employed version 19 of the Suzaku Software included in the 
 \textsc{HEAsoft} software package and followed the procedures described in the 
Suzaku ABC Guide.\footnote{heasarc.gsfc.nasa.gov/docs/suzaku/analysis/abc/ .} 
Following standard practices, 
we excluded times within 436 s of \emph{Suzaku} passing through the South Atlantic Anomaly 
and we also excluded the data when the line of sight was elevated above the Earth's limb by less than $5^\circ$, 
or was less than $20^\circ$ from the bright-Earth terminator. 
Moreover, we excluded time windows during which the spacecraft was passing through the 
low cut-off rigidity of below 6 GV. 

For the XIS we considered only events with GRADE = 0,2--4,6 and removed hot and flickering pixels 
using \textsc{sisclean}; for the spectral analysis, the response matrices were generated  
with \textsc{xisrmfgen} and using ray-tracing simulations with \textsc{xissimarfgen}. 
With all the aforementioned data selection criteria applied, the resulting total effective exposure
is roughly 136~ks for each XIS. 
The XIS events of IGR\,J16479--4514 were accumulated in each of the three XIS cameras 
within a circular region (3 arcmin radius) centred on the target, while the backgrounds 
were estimated from an annulus with radii 5 and 7 arcmin.
During the observation, the standard criterion for the attitude determination requiring that the residuals are
less than $0.005^\circ$  could not be fullfilled, with residuals still remaining at
about $0.008^\circ$ for about 20 hours, in the temporal window from February 25 01:00
through 22:00, but still smaller 
than the \suz\ point spread function (PSF) for usual point-like source analysis.

In the HXD-PIN (no significant emission was detected in the GSO) 
the net exposure time, after dead-time correction (live time: 92.8\%) is 141.9 ks. 
To subtract non-X-ray background (NXB) events that are included in the HXD-PIN 
we used the `tuned' NXB synthesized background \citep{Fukazawa2009}.
After subtracting the synthesized NXB from the HXD-PIN data, 
we also subtracted contributions of the cosmic X-ray background (CXB) 
using the model reported in \citet{Gruber1999}
and of the Galactic Ridge X--ray emission (GRXE; \citealt{Valinia1998}).

To ensure applicability of the \chisq\ statistics, the
net XIS spectra were rebinned such that at least 20 counts per
bin were present.
All spectral
uncertainties and upper-limits are given at 90\% confidence for
one interesting parameter.
Data were analysed using {\sc ftools} version 6.11.1  and {\sc xspec} version 12.
In the spectral fitting we used the photoelectric absorption
model {\sc phabs} in {\sc xspec} with the interstellar
abundances of \citet{Wilms2000}
and  cross section table set of \citet{bcmc1992}.
The three joint XIS0, XIS1 and XIS3 spectra were fitted together,
including constant factors to allow for normalization uncertainties between the instruments.
For the timing analysis photon arrival times were corrected to the Solar System
barycenter.

  	\section{Analysis and Results\label{result}}

\subsection{Light curves}
\label{sec:lc}

\begin{figure*}
\begin{center}
\centerline{\includegraphics[width=12.5cm,angle=-90]{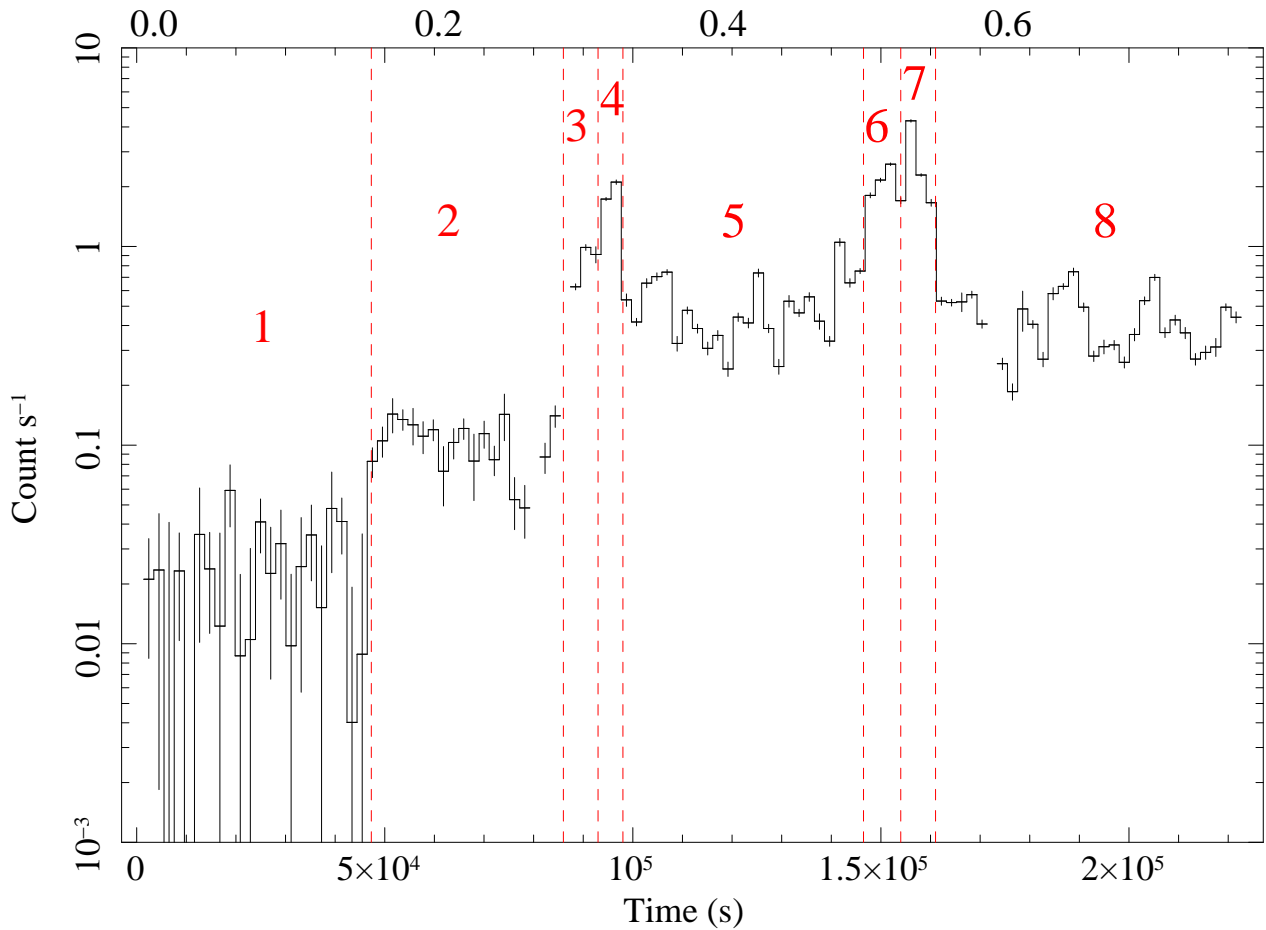}} 
\caption{Background-subtracted light curve of \src\ 
in the 0.2--10 keV energy range, obtained 
combining net source count rates observed by the three XIS units. 
In the upper x-axis, the
orbital phase is indicated, assuming the orbital period P$_{orb}$=286792~s    (Romano et al. 2009),
assuming epoch 54547.05418 MJD as orbital phase $\phi$=0 (Bozzo et al. 2009).
The bin size is 2048 s. Vertical dashed lines and numbers 
indicate the intervals of the eight time-selected spectra 
reported in Table~\ref{tab:spec}. }
\label{fig:lc}
\end{center}
\end{figure*}

\begin{figure*}
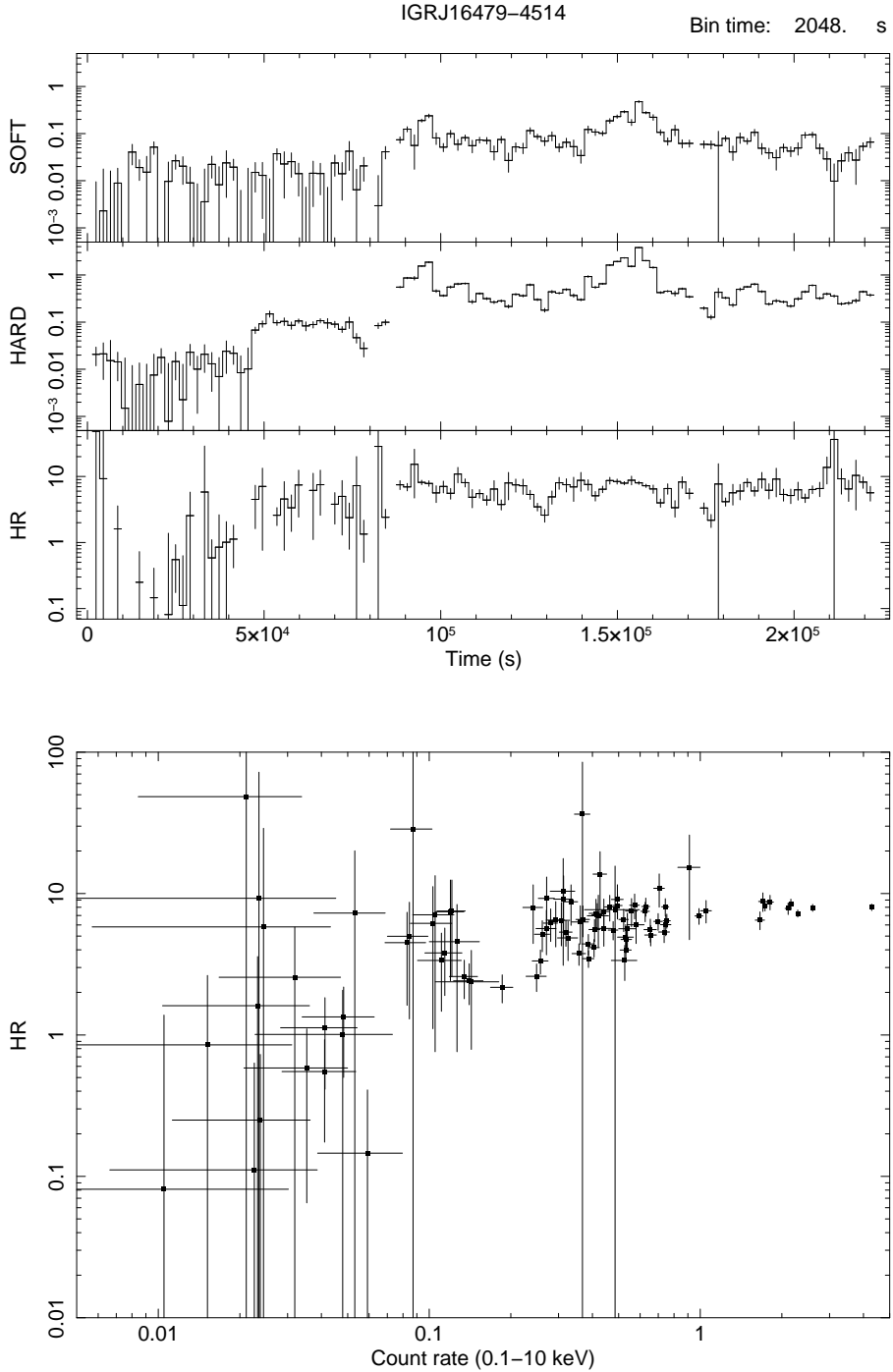

\begin{center}
\begin{tabular}{cc}
\includegraphics[height=12.5cm,angle=-90]{fig2a.ps} \\
\includegraphics[height=12.5cm,angle=-90]{fig2b.ps}
\end{tabular}
\end{center}
\caption{Hardness ratio (hard/soft) variations with time ({\em Upper panel}) and 
with the source intensity ({\em Lower panel}).
The bin size is 2048 s. Soft and hard energy ranges are below and above 3 keV, respectively.
} 
\label{fig:hr}
\end{figure*}

The XIS  barycentred and background-subtracted light curve of \src\ in the energy range
0.2--10 keV,
is shown in Fig.~\ref{fig:lc}, while the hardness ratio (HR) between net counts
extracted in the hard (above 3 keV) and the soft (below 3 keV) energy bands
is reported in Fig.~\ref{fig:hr}, both versus time ({\em upper panel}) and in dependence of
the source intensity ({\em lower panel}).
Numbers in Fig.~\ref{fig:lc} indicate time intervals which show different 
source intensity behaviour. 
At the beginning of the observation a very low intensity state is present (n.~1),
consistent with being the X--ray eclipse produced by the companion star.
Indeed, folding the  light curve on the refined 
orbital period P$_{orb}$=286792$\pm{43}$~s  \citep{Romano2009:sfxts_paperV}
and assuming the epoch 54547.05418 (MJD) as orbital phase $\phi$=0 \citep{Bozzo2009}, we obtain
the orbital phases reported as top x-axis in Fig.~\ref{fig:lc}.  
The uncertainty on the epoch of $\phi$=0  can be estimated in about $\phi$=$\pm{0.065}$, obtained
extrapolating the error on the orbital period to the time of the \suz\ observation (432 orbital
cycles between the epoch 54547.05418 MJD and the \suz\ observations).

The \suz\ observation did not cover the ingress time, so we are unable to determine the exact eclipse duration.
We can at least constrain it to be between 46~ks (the duration of the eclipse at hard X--rays) 
and 143~ks (that is, 0.5 in phase, assuming a perfectly edge-on system with a tight orbit). 
An eclipse duration of about 0.6~days (52~ks)
as proposed by \citealt{Jain2009:16479} could imply that we might be seeing almost the full duration of
the eclipse.
After the eclipse (time interval n.~2) an intermediate level of emission is observed, lasting about 38~ks,
during which
the soft X--ray intensity below 3 keV is indistinguishable from  
the emission during eclipse (Fig.~\ref{fig:hr}),
while at harder energies the source appears uneclipsed, although with a reduced intensity 
with respect to the uneclipsed emission displayed in the intervals n.~5 and n.~8.

Two flares are present peaking at 2~counts~s$^{-1}$ and 4~counts~s$^{-1}$
in the time intervals n.~4 and n.~7, respectively. 
Intervals n.~3 and n.~6 can be considered the rise to these two flare peaks, while
intervals n.~5 and n.~8 are characterized by a very variable, intra-flare emission,
which covers about  40\% of the orbital cycle.

\begin{figure}
\begin{center}
\centerline{\includegraphics[width=6.5cm,angle=-90]{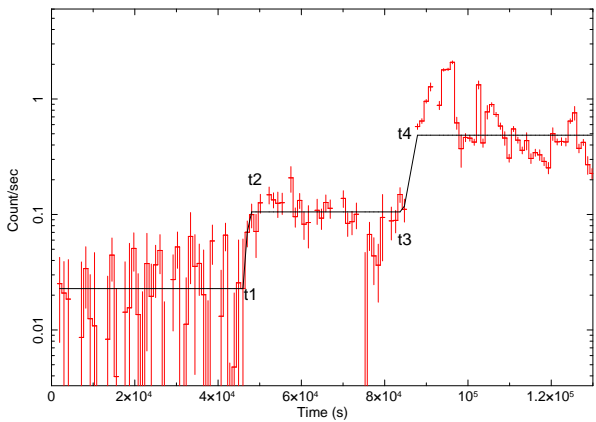}}
\caption{Double ramp-and-step function fitting the initial part of the XIS light curve.
 The four times  discussed in the Sect.~\ref{sec:lc} are indicated.}
\label{fig:ramp}
\end{center}
\end{figure}

The X--ray light curve shows two sharp transitions from the eclipse towards the uneclipsed X--ray emission.
To better estimate the time of the eclipse, we simply modeled the light curve 
with two ramp-and-step functions, characterized by seven parameters:
the count rates of the three ``steps'' in the light curve and the four times, t${_i}$, 
displayed in Fig.~\ref{fig:ramp}.
The fit resulted in the following times measured from the beginning of the observation: 
t$_1$=$46 \pm{1}$~ks, 
t$_2$=$47 \pm{1}$~ks,  
t$_3$=$85 \pm{2}$~ks, 
t$_4$=$88 \pm{2}$~ks.
The time t$_1$
corresponds to the date 55981.5131~$\pm{0.0116}$ (MJD).

\subsection{Timing Analysis}   
\label{sec:timing}

Barycentred light curves with 8 seconds bin time were produced for each unit
of the X-ray Imaging Spectrometer (XIS0, XIS1, XIS3) 
in three energy bands (0.2--3,  3--10, 0.2--10 keV).  
We considered  two time  ranges:  i) 
the entire duration of the \suz\ observation (interval
from 1 to 8 in Fig.~\ref{fig:lc}) and 
ii) the part of the observation excluding the eclipse and the transition to the 
uneclipsed emission (time intervals from 3 to 8 in
Fig.~\ref{fig:lc}). 
In order to search for spin periodicities  in the  XIS light curves, 
we used the  Lomb-Scargle periodogram method by means of the fast
implementation of Press \& Rybicki (1989) and Scargle (1982), which is
generally  
preferred for data set with  gaps and unequal sampling. Periodicities were
searched in the frequency range  from i) 0.000013 Hz  or ii) 
0.000022 Hz (after which the sensitivity is reduced due to the finite length
of the light curves)  to 0.0625 Hz  (corresponding to the Nyquist  
frequency  of the data set).  No significant and unambiguous  evidence for
coherent modulation was found in the periodograms.

\subsection{Spectroscopy}
\label{sec:spec}

\begin{figure}
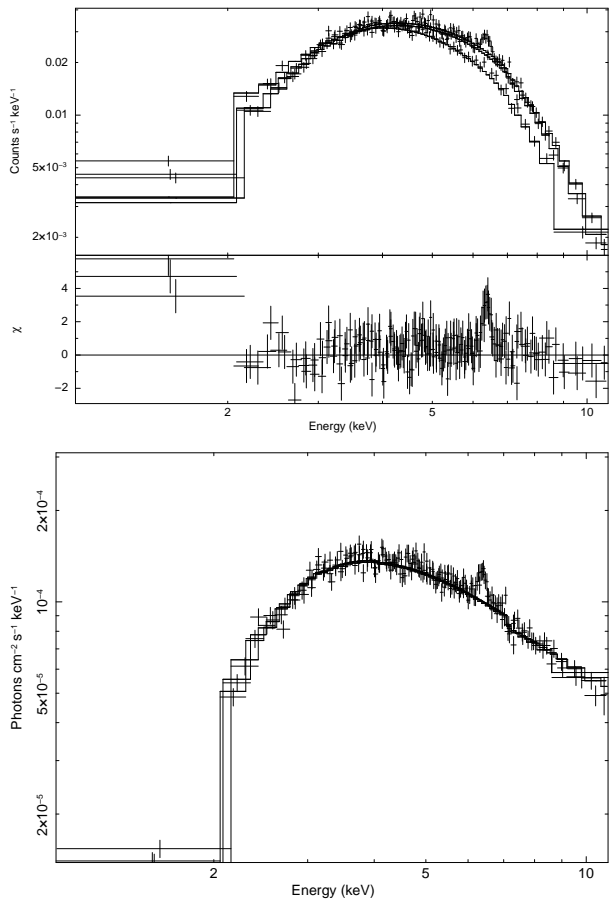

\begin{center}
\begin{tabular}{cc}
\includegraphics[height=8.0cm,angle=-90]{fig4a.ps} \\
\includegraphics[height=8.0cm,angle=-90]{fig4b.ps}
\end{tabular}
\end{center}
\caption{XIS spectrum extracted from the whole \suz\ observation. 
Count spectrum is shown ({\em Upper panel}), together
with the residuals (in units of standard deviations) of the data to
the absorbed power-law model (see Table~\ref{tab:av_spec}, first column, for the spectral
parameters). {\em Lower panel} shows the photon spectrum, graphically rebinned
to better show the residuals around 6.4~keV.
} 
\label{fig:av_spec}
\end{figure}

We first analysed the joint XIS spectra extracted from the entire observation,
corresponding to a net integration time of 136~ks.
The net source count rates (0.2--10 keV) in the three XIS were the following:
0.157$\pm{0.001}$~count~s$^{-1}$ (XIS0), 
0.144$\pm{0.001}$~count~s$^{-1}$ (XIS1) and  
0.165$\pm{0.001}$~count~s$^{-1}$ (XIS3).
Adopting an absorbed power-law model resulted in a flat spectrum (photon index, $\Gamma$, of 1.35), 
a strong absorption (N$_{\rm H}$$\sim$10$^{23}$~cm$^{-2}$) 
and in  positive residuals around  6.4~keV (Fig.~\ref{fig:av_spec}).
Adding a narrow Gaussian line to the power-law continuum, we obtain a better fit, resulting 
in a line energy of $6.37 \pm{0.03}$~keV, compatible with being produced by Fe~K${_\alpha}$ fluorescence.
The spectral parameters of the average X--ray spectrum are listed in Table~\ref{tab:av_spec}.
The average flux, F, of 1.16$\times$10$^{-11}$~erg~cm$^{-2}$~s$^{-1}$ (corrected for the absorption), 
translates into an X--ray luminosity 
of 10$^{34}$~erg~s$^{-1}$ (assuming a distance of 2.8~kpc, \citealt{Nespoli2008}).
Note however that, given the large uncertainty in the distance determination 
\citep{Nespoli2008}, 
this luminosity  can range from 1.7$\times$10$^{33}$~erg~s$^{-1}$ (at 1.1~kpc) to
8.2$\times$10$^{34}$~erg~s$^{-1}$ (at 7.7~kpc).

The hard X-ray spectrum extracted from the HXD-PIN field of view (FOV) is much brighter than 
the extrapolation at higher energies of the XIS power-law best-fit. 
Indeed, the  HXD-PIN flux ($\sim$1.5$\times$10$^{-10}$~erg~cm$^{-2}$~s$^{-1}$, after taking into account
CXB and GRXE) in the energy range 
15--25 keV is $\sim$15 times greater than that calculated extrapolating at high energies 
the power-law best fit to the XIS spectrum. Note however that this latter extrapolated flux
should be considered as a conservative upper limit, since \src\ typically shows a spectral 
cut-off above 10 keV. 
This implies a strong contamination by  
other X--ray sources in the FOV, probably a residual contamination by the 
Z-track low-mass X-ray binary GX340+0, located at about 34 arcmin
from \src, and/or an unknown bright hard X--ray transient.
For this reason, we will not discuss it further.

\begin{table}
\begin{center}
\caption[]{Spectral results of the time averaged spectrum from the \suz/XIS observation.
$\Gamma$ is the power-law photon index. A Gaussian line is required at $\sim$6.4 keV (see Fig.~\ref{fig:av_spec}). Flux is in
the 1--10~keV energy range in units of
10$^{-11}$~erg~cm$^{-2}$~s$^{-1}$ and is corrected for the
absorption, N$_{\rm H}$ (in units of $10^{22}$~cm$^{-2}$). 
The flux of the iron line, F$_{line}$, is in units of  10$^{-6}$~ph~cm$^{-2}$~s$^{-1}$. 
L$_{\rm X}$  is the X--ray luminosity (1--10 keV), in units of  10$^{34}$~erg~s$^{-1}$, for an assumed distance of 2.8~kpc.   }
\begin{tabular}{llll}
 \hline
\hline
\noalign {\smallskip}
Parameter   &           Power-law       &     Power-law + Gaussian line                \\
\hline
\noalign {\smallskip}
N$_{\rm H}$             &  $ 9.5 \pm{0.3}$             &      $ 9.5 \pm{0.3}$             \\
$\Gamma$                &  $1.33 \pm{0.04}$           &     $1.35 \pm{0.04}$              \\
F$_{line}$      &        $-$              &        7.4 $\pm {1.5}$         \\
EW (eV)                                                 &         $-$                 &         70$\pm {10}$           \\
E$_{line}$ (keV)                                        &         $-$                 &      $6.37  \pm{0.03}$             \\
Unabs. Flux             &    1.16                    &     1.16                       \\
L$_{\rm X}$             &    1.09               &        1.09                        \\
$\chi^{2}_{\nu}$/dof    &   1.018/ 2172                        &          0.991/2169         \\
\noalign {\smallskip}
\hline
\label{tab:av_spec}
\end{tabular}
\end{center}
\end{table}

The hardness ratio is variable (Fig.~\ref{fig:hr}) especially during eclipse, where the emission
appears  softer and/or less absorbed, so we next
explored the variability of the spectral parameters in dependence of the orbital phase
performing a time-selected spectroscopy extracting counts from the same intervals indicated in Fig.~\ref{fig:lc}.
During the eclipse a remaining detectable flux is present, as typical in eclipsing high mass X--ray binaries,
mainly produced by Thomson scattering into the line of sight of the direct component
by free electrons residing in the supergiant wind  \citep{Haberl1991}.
We fit the spectra with an absorbed power-law model, resulting in a less absorbed and softer spectrum
during the eclipse compared to the other time intervals (Table~\ref{tab:spec}).
The absorption during the eclipse is consistent with the interstellar reddening
observed to the optical counterpart \citep{Nespoli2008}, while that measured outside the eclipse is significantly higher,
and consistent, within the uncertainties, with being constant, at about N$_{\rm H}$$\sim$10$^{23}$~cm$^{-2}$.
The X--ray emission is harder when the source is brighter (X--ray flares), a typical
behaviour observed in SFXTs and in accreting pulsars.

\begin{figure*}
\begin{center}
\centerline{\includegraphics[height=17.5cm,width=18.0cm,angle=-270]{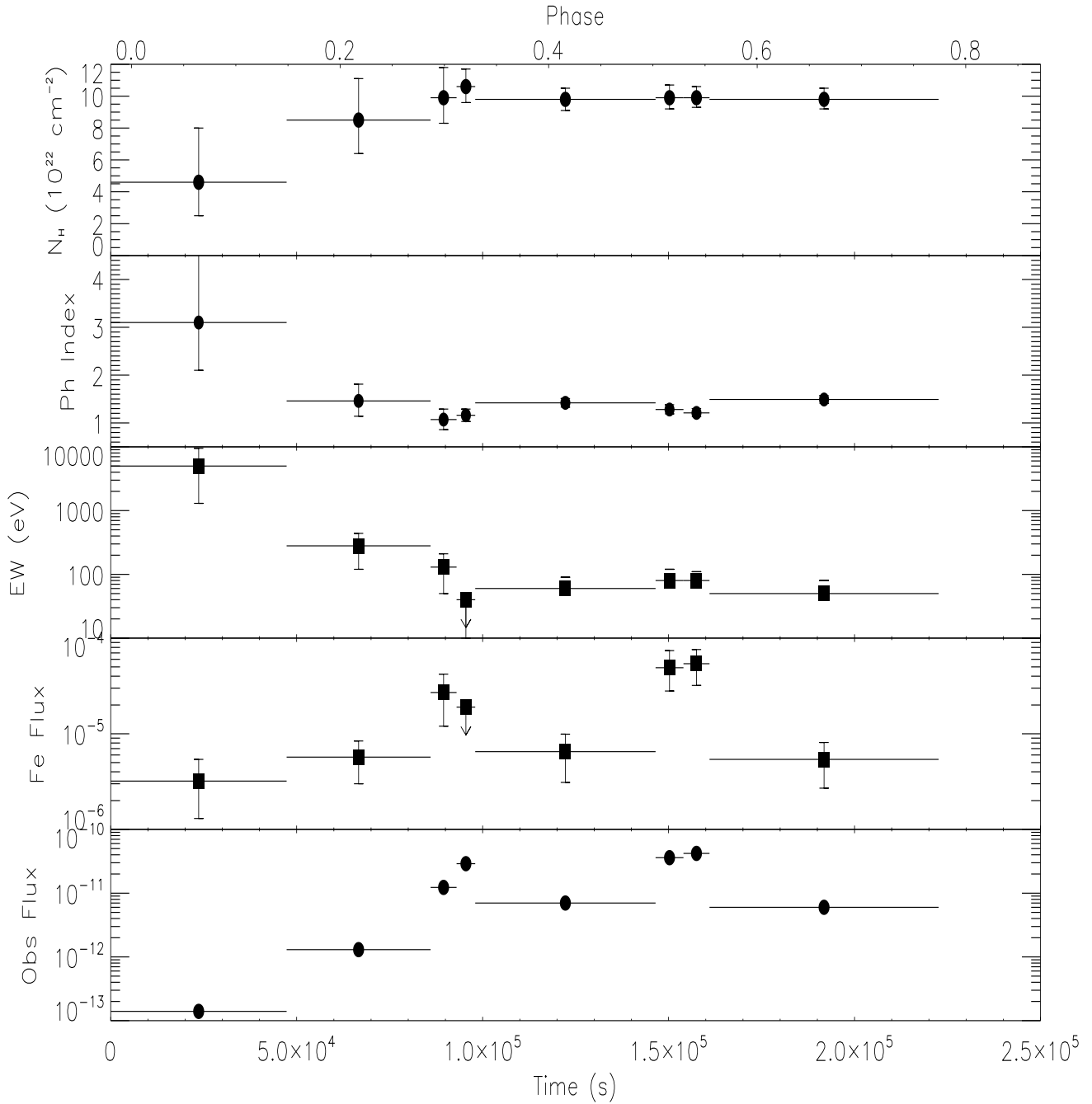}}
\caption{Variability of the best-fit spectral parameters adopting an absorbed power-law
model together with a Gaussian line at 6.4~keV (see Table~\ref{tab:spec}). 
The observed (not corrected for the absorption) flux is in the energy range 1--10 keV and in units of erg~cm$^{-2}$~s$^{-1}$.
The flux of the iron line is in units of photons~cm$^{-2}$~s$^{-1}$.
The orbital phase is indicated in the upper x-axis (see text).}
\label{fig:timesel_pow}
\end{center}
\end{figure*}

\begin{figure}
\begin{center}
\centerline{\includegraphics[width=6.5cm,angle=-270]{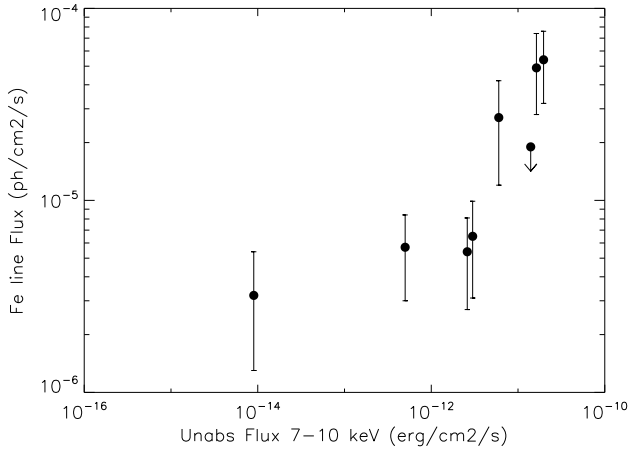}}
\caption{Fe K$_{\alpha}$ emission line flux versus the flux in the energy range 7--10 keV, corrected for the absorption. 
}
\label{fig:timesel_fe}
\end{center}
\end{figure}

We next added a narrow Gaussian line to the single power-law model, resulting in the parameters reported in Table~\ref{tab:spec} and 
shown in Fig.~\ref{fig:timesel_pow}. The resulting continuum power-law parameters are fully compatible, within the
uncertainties, with those derived without the Gaussian line, so we do not report them again in Table~\ref{tab:spec}.
The energy of the narrow line indicates K$_{\alpha}$ emission from neutral (or almost neutral) iron.
Its equivalent width (EW) has been calculated with respect to the power-law continuum, and its large value
during the eclipse is due to the fact that it has been calculated with respect to the
scattered component. The direct eclipsed X--ray emission is unknown, given the large source variability. 
The intensity of the fluorescence Fe~K$_{\alpha}$ line is correlated with the unabsorbed hard X--ray flux (above 7 keV),
at least outside the eclipse (Fig.~\ref{fig:timesel_fe}).

Finally, we adopted a power-law model together with a Gaussian line, but now fixing 
the power-law photon index to 1.35, the value obtained from the average spectrum. 
The fits are acceptable and the resulting values of the absorbing column density show now some
variability,  again  with the less absorbed spectrum seen during the eclipse.
More complex models are not required by the count statistics 
of the data and result in unconstrained spectral parameters.

\begin{table*}
\begin{center}
\caption[]{$Suzaku$/XIS results of the  time-selected spectroscopy (numbers
mark the same time intervals displayed in Fig.~1) with an absorbed
power-law model. Fluxes are 
in units of 10$^{-11}$~erg~cm$^{-2}$~s$^{-1}$. Both observed (not corrected for the absorption)
and unabsorbed (corrected for the absorption) fluxes are reported. 
L$_{\rm X}$  is the X--ray luminosity (1--10 keV), in units of  10$^{34}$~erg~s$^{-1}$, for an assumed distance of 2.8~kpc. 
The absorbing column density, N$_{\rm H}$, is in units of
$10^{22}$~cm$^{-2}$. 
}
\begin{tabular}{lllllllll}
 \hline
\hline
\noalign {\smallskip}
                           &           1                          &     2        &    3        &      4         &        5          &   6    &   7 &  8    \\
\hline
\noalign {\smallskip}
Power-law   &   &  &   &  &  &  &   &  \\
\hline
\noalign {\smallskip}
N$_{\rm H}$            &  $ 4.6 ^{+3.4} _{-2.1}$        &    $ 8.5 ^{+2.6} _{-2.1}$         &  $9.9 ^{+1.9} _{-1.6}$  &  $10.6 ^{+1.1} _{-1.0}$    & $9.8 ^{+0.7} _{-0.7}$  &  $9.9 ^{+0.8} _{-0.7}$   &    $9.9 ^{+0.7} _{-0.6}$        &   $9.8 ^{+0.7} _{-0.6}$   \\

$\Gamma$               &  $3.1 ^{+1.5} _{-1.0}$   &    $1.46 ^{+0.35} _{-0.32}$  &  $1.07 ^{+0.22} _{-0.21}$  &  $1.16 ^{+0.13} _{-0.13}$  & $1.42^{+0.09}_{-0.09}$ &  $1.28^{+0.10}_{-0.09}$  &     $1.21^{+0.09}_{-0.08}$         &     $1.49 ^{+0.08} _{-0.08}$   \\
Obs. Flux (1-10 keV)            &    0.014                  &     0.13            &  1.23                   &  2.9                      & 0.7                    &  3.6                  &  4.2        &    0.6  \\
Unabs. Flux  (1-10 keV)           &    0.045                 &     0.22           &  1.9                   &  4.7                     & 1.2                    &  5.9                     &  6.8         &    1.1  \\
L$_{\rm X}$     &    0.042     &    0.21        &    1.8       &      4.4         &     1.1      &     5.5        &      6.4       &     1.0      \\
$\chi^{2}_{\nu}$/dof   &     1.672/48            &     1.047/64         &  0.836/88              &  1.092/227               &     0.922/634           &  1.001/435 &      0.887/518   &  1.130/656  \\
\noalign {\smallskip}
\hline
Power-law + Gaussian line   &   &  &   &  &  &  &   &  \\
\hline
\noalign {\smallskip}
F$_{line}$ (10$^{-5}$~ph~cm$^{-2}$~s$^{-1}$)    &   $0.32 ^{+0.22} _{-0.19}$          &   $0.57 \pm{0.27}$    & $2.7 ^{+1.5} _{-1.5}$     &   $<$1.9     &   $0.65 ^{+0.34} _{-0.34}$    &   $4.9 ^{+2.5} _{-2.1}$     &   $5.4 ^{+2.2} _{-2.2}$      &    $0.54 ^{+0.27} _{-0.27}$    \\

EW (eV)                                         &    $5500 ^{+4500} _{-3700} $ $^a$        &  280 $\pm{160}$    &  130$\pm{80}$     &    $<$40    &   60$\pm{30}$     &  80$\pm{40}$      &   80$\pm{30}$      &    50$\pm{30}$    \\
E$_{line}$ (keV)                                &   $6.57 ^{+0.22} _{-0.26} $          &  $6.33\pm{0.20}$     &   $6.41 ^{+0.08} _{-0.07} $    &  6.4 (fixed)      &   6.38$\pm{0.06}$    &   6.34$\pm{0.06}$     &    $6.37 ^{+0.04} _{-0.05} $    &     6.32$\pm{0.09}$      \\
$\chi^{2}_{\nu}$/dof                              &   1.576/45          &  0.934/61     &   0.771/85    &  1.097/226      & 0.910/631      &  0.973/432     &  0.861/515      &  1.118/653     \\
Unabs. Flux (7--10 keV)         &  0.0009 & 0.05 & 0.60  & 1.40 & 0.30  & 1.63  & 1.97  & 0.26  \\
\noalign {\smallskip}
\hline
Power-law + Gaussian line   &   &  &   &  &  &  &   &  \\
\hline
\noalign {\smallskip}
N$_{\rm H}$            &  $ 1.3 ^{+0.6} _{-0.5}$        &    $ 7.3 ^{+1.3} _{-1.1}$         &  $12 \pm{1}$  &  $11.9 \pm{0.6}$    & $9.2  \pm{0.3}$  &  $10.2  \pm{0.4}$   &   $10.6  \pm{0.4}$    &   $8.8 \pm{0.2}$   \\
$\Gamma$               &   1.35 fixed    &   1.35 fixed   &   1.35 fixed   &   1.35 fixed   &   1.35 fixed   &   1.35 fixed  &   1.35 fixed  &   1.35 fixed \\
$\chi^{2}_{\nu}$/dof   &     1.920/46            &     0.923/62         &  0.799/86              &  1.118/227               &     0.913/632          &  0.963/433 &      0.870/516   &  1.132/654  \\
\hline
\noalign {\smallskip}
\label{tab:spec}
\end{tabular}
\end{center}
\begin{small}
$^{a}$ The large EW during eclipse is calculated with respect to the scattered power-law continuum. \\
\end{small}
\end{table*}

	\section{Discussion \label{discussion}}

The long \suz\ observation we have reported on here, allows us to investigate
the properties of a SFXT as continuously as possible 
within one single orbital cycle for the first time. 
This has been made possible by the short orbital period of \src\, which
is the SFXT with the shortest orbit known to date.
The X--ray source shines with an average low luminosity L$_{\rm X}$$\sim$10$^{34}$~erg~s$^{-1}$, 
(if a distance of 2.8 kpc is assumed) and no bright flares 
(exceeding L$_{\rm X}$$\sim$10$^{36}$~erg~s$^{-1}$)
have been caught during this observation.
Nevertheless, a high amplitude variability is observed,
spanning more than two orders of magnitude (including the X--ray eclipse).
The absence of signatures from highly ionizing illuminating continuum further confirms
a low source X--ray luminosity.

We can distinguish several types of variability on different timescales, especially
evident in the energy range 3--10 keV.
On long timescales of tens of ks we observe three main intensity states: 
the end of the eclipse
by the supergiant companion (which we covered for about 46 ks), 
a transition phase to the uneclipsed emission (time interval n.2 in Fig.~1, 
lasting 38 ks) and
the intra-flare low luminosity emission which covers 
most of the remaining orbit (40\%), lasting around 110 ks (time intervals n.5 and n.8).


X--ray flares with different durations are present: two main low 
luminosity flares, lasting 10--15 ks, 
with a complex and structured shape, punctuate the orbit, 
reaching peak luminosities
of 4.4 and 6.4$\times$10$^{34}$~erg~s$^{-1}$ (at 2.8 kpc), separated by about 
0.2 in orbital phase.
Their 0.2--10 keV spectra are harder than the intra-flare X--ray emission,
as usually observed in SFXTs and accreting pulsars.
Even in the part of the orbit in-between the two main low luminosity flares,
the source exhibits a large intensity variability on timescale of 1000 s.
This low luminosity flaring activity had already been unveiled
in SFXTs (and in \src) thanks to
\sw/XRT long term monitoring observations \citep{Sidoli2008:sfxts_paperI}.


In the softer light curve the X--ray eclipse appears longer and there
is no evidence for the presence of the transition phase (time interval n.2)
clearly present in the hard light curve.
Thus the eclipse egress appears to be energy dependent.
This behaviour is similar to that reported by \citet{Jain2009:16479}
from the shape of the \src\ folded light curves
observed with \xte\ and with \sw/BAT, where the X--ray eclipse
seems more evident at higher energies and with sharper transitions.
There is also a similarity of the \src\ temporal variability near eclipse egress
with what has been observed in the HMXB 4U1700--37 (\citealt{vanderMeer2005} and references therein), 
showing a variable eclipse duration and a similar egress phase.
The long duration (38~ks) of this phase roughly corresponds to a region 
as large as the companion radius (although a precise determination
of both the donor radius and of the system inclination is lacking), 
probing the innermost structure 
of the supergiant wind. 

On shorter timescales of around 1000~s we observe two sharp transitions, indicated
in Fig.\ref{fig:ramp} by the four contacts t$_i$:
(1) from the eclipse to time interval n.~2, which shows an intermediate level
between the eclipse remaining flux and the un-eclipsed flux, and
(2) from time interval n.~2 to the first flare.
Given that the compact object is point-like with respect to the supergiant companion,
the first short transition likely probes the transition out off behind the stellar 
atmosphere, which, assuming an orbital velocity of a few hundreds km~s$^{-1}$, results
in a linear size of around a few 10$^{10}$~cm.
The second sharp transition is interestingly coincident with the rising phase to
the first X--ray flare, suggesting a possible physical connection (see the discussion below).


An iron line is detected along the orbit, with a line energy of 6.4 keV, 
indicative of neutral iron or an ionization state lower
than Fe~{\sc xviii} \citep{Kallman1982}.
A correlation is observed between the iron  fluorescent line flux and the 
illuminating unabsorbed flux above 7~keV,  as expected \citep{Inoue1985} outside the eclipse  (Fig.~\ref{fig:timesel_fe}).
The line flux observed during eclipse does not show evidence 
for a  reduction, within the  large uncertainties, with respect to the value detected
during the intra-flare emission, so  most of the 
fluorescing matter is located far away from the compact object.
Note that we calculate the equivalent width of the line with respect to a single
power-law continuum. 
If the fluorescing matter is optically thin to the illuminating continuum,
the luminosity of the iron K$_{\alpha}$ line depends on the spectrum
and X--ray luminosity of the central source, on the covering fraction of the illuminated
matter and iron abundance (e.g. \citealt{Sako1999}). 
Given the large uncertainties on the EW (especially huge during the eclipse phase) and  
the variability of the covering factor along the orbital phase,
also considering the likely presence of
different large-scale structures in the supergiant wind which can be obvious, although unknown, 
sites of fluorescing material, it is not possible to gain further constraints on the properties of the 
reprocessing wind (also the supergiant surface becomes a reprocessor only visible at certain orbital phases).

Considering the spectroscopy,
a more physical description of the X--ray spectrum in an eclipsing HMXB involves 
the presence of two power-law continua, accounting for  the direct and scattered component
with the latter produced by Thomson scattering by free electron in the companion wind and/or on the supergiant
surface \citep{Haberl1991}. Both the direct and the scattered components 
should have the same spectral slope, since
Thomson scattering is energy independent. This implies that the only way to disentangle 
these two components is the different absorptions, the scattered component being 
less absorbed than the direct one. During the X--ray eclipse, where the direct component is unseen,
the X--ray emission should only be due to scattering. Indeed, fitting this
spectrum with a power-law with a photon index fixed at the average slope of $\Gamma$=1.35,
the eclipse shows the lowest absorption, consistent with the interstellar value \citep{Nespoli2008}.
Unfortunately, given the high absorption (local and interstellar) and the low statistics, 
we could not find better contraints to the
spectral parameters adopting more complex models.

Also the egress X--ray emission (n.~2) in \src\ is likely dominated by the scattered component, given the
high EW of the iron line compared to that measured during the out-of-eclipse emission (time
intervals from n.~3 to n.~8). 
A possibility is that most of the direct X--ray emission in this orbital region is blocked by a 
large scale wind structure, as that invoked to explain the periodically recurrent outbursts
in the SFXT IGR~J11215--5952 \citep{Sidoli2007}, or by dense shells or other kind of inhomogeneities
present in hot massive stars winds (see, e.g., \citealt{Oskinova2012} or \citealt{Lobel2008}).
This implies that  most of the X--ray emission in this time interval is scattered X--ray emission, which is less
absorbed because 
mainly produced by less dense wind material
located off the orbital plane and farther away from
the companion star.
This could explain why the observed column density in time interval n.~2 
is not larger than in the remaining (fully out-of-eclipse) part of the orbit. 

The likely presence of large scale structures in the supergiant wind is also suggested
by the orbital phase of the first flare, which is compatible (within the uncertainties, see Sect.~\ref{sec:lc})
with the location along the orbit of other bright flares observed in the past in \src\ ($\phi$$\sim$0.36; \citealt{Bozzo2009}).
This suggests that the flares are triggered by a higher accretion rate during the passage inside
a phase-locked denser wind component, like the one predicted by \citet{Sidoli2007}.
This seems to be also indicated by the fact that, interestingly, the egress X--ray emission (n.~2) ends 
with the sharp transition 
to the first X--ray flare (see Sect.~\ref{sec:lc}). 
If this explanation is correct, after the first flare, 
the compact object lies  in-between the observer
and the dense wind structure, with the direct X--ray emission finally dominating 
the observed X--rays along the remaining part of the orbit. 

The second flare, spaced by about 0.2 in phase, could be reconciled by the crossing of a similar
wind component, if the orbit is eccentric (as in the light curves
simulated by \citealt{Ducci2009}), or by the compact object approaching the periastron passage. 
Unfortunately, we lack any information about the other orbital parameters (e.g. eccentricity)
to be able to confirm this hypothesis.


The absorbing column density is always in excess of the interstellar reddening and does not show evidence
for variability along the orbit (if also the slope of the power-law is mantained free), except that
during the eclipse.
On the other hand, if the power-law slope is fixed to the value of the average spectrum ($\Gamma$=1.35),
variability is present, with more absorption during the low luminosity flares.

The only known orbital parameter is the period of 3.32~days. 
We did not observe the eclipse ingress, so we cannot constrain the companion radius, R$_{opt}$.
On the other hand, R$_{opt}$ can be estimated from the supergiant spectral type and requiring that
the star does not overflow its Roche lobe. 
The spectral type of the supergiant (O8.5I or O9.5I) suggests a mass in the range M$_{opt}$=30--34~$\msun$
and a radius of 22-23~R$_{\odot}$ \citep{Martins2005}.
A further constraint on the stellar size comes from the fact that the companion should 
not overlflow its Roche lobe radius \citep{Eggleton1983}. This 
suggests a donor mass around M$_{opt}$=35~$\msun$ and a stellar radius R$_{opt}$=20~R$_{\odot}$. 
Adopting these values, the orbital period translates into an orbital separation
of about 2.2$\times$10$^{12}$~cm. 

We can use the X--ray eclipse to probe the supergiant properties, as follows (see, e.g., \citealt{Lewis1992}).
The average total intensity during the eclipse is about 5\% of the
out-of-eclipse, intra-flare X--ray intensity, so the wind density  can be obtained
as $n_{\rm w}$ = $0.05 / (a \sigma_{\rm T})$, where  $\sigma_{\rm T}$ is the Thomson scattering cross section
and {\em a} is the orbital separation, which we 
assume as the characteristic path lenght through the system.
This results in a wind density at the orbital separation, $\rho$$_{\rm w}(a)$, of 7$\times$10$^{-14}$~g~cm$^{-3}$.
The mass continuity equation can be used to derive the ratio between the wind mass-loss rate
and the terminal velocity, as $\dot{M}_{w}/ v_{\infty}$=4~$\pi$~$a~(a-R_{opt}$)~$\rho_{\rm w}(a)$,
assuming a spherical geometry for the outflowing wind and a velocity 
gradient for the wind velocity, $\beta$, of 1 \citep{Castor1975}.
This results in a ratio  $\dot{M}_{w}/ v_{\infty}$=7$\times$10$^{-17}$~M$_{\odot}$/km.
For  terminal velocities in the range 500--3000~km~s$^{-1}$, the mass loss rate is in the range 
$\dot{M}_{w}$=1--7$\times$10$^{-6}$~M$_{\odot}$/yr.
Assuming direct accretion from the wind, the accretion rate can be estimated as 
$\dot{M}_{\rm acc}$=$(\pi R_{acc}^{2} / 4\pi a^{2})$$\cdot$$\dot{M}_{w}$, where R$_{acc}$ is the accretion radius. 
This accretion rate translates into an X--ray luminosity L$_{\rm X}$=3--15$\times$10$^{36}$~erg~s$^{-1}$
for the range of parameters estimated before, which is two orders of magnitude greater than that we observe.

Given the high wind density we have calculated, it is unlikely that the  low luminosity in \src\ is
due to the direct wind accretion. 
Our findings seems to agree with the recent
results by Oskinova et al. (2012), who have demonstrated
that simple Bondi-Hoyle accretion from a clumpy wind over-predicts the observed X--ray variability in HMXBs.
This suggests the presence of some mechanism able to damp 
the strong X--ray variations implied by the structured supergiant winds
\citep{Oskinova2012}.
This can be due to the details (still poorly known) 
of the interaction of the accretion flow with the shocks in the accretion wake, leading, for example, to a 
transitional case of accretion regime, intermediate between Roche Lobe Overflow and direct accretion (as originally proposed for
the SFXT IGR~J16418--4532, given its short orbital period \citep{Sidoli2012}, which displays a similar X--ray intensity variability).

The role of the magnetospheric surface in reducing the accretion rate onto the NS in SFXTs have been
studied by several authors (\citealt{Bozzo2008}, \citealt{Ducci2010}, \citealt{Shakura2012}).
Bozzo et al. (2008) have invoked magnetar-like NS with slow pulsations, to halt the accretion most of the time 
and allow only a residual flow of matter (producing $\sim$10$^{34}$~erg~s$^{-1}$) 
by means of Kelvin-Helmholtz instability. 
\citet{Shakura2012b} have proposed that the low X--ray luminosity 
state ($\sim$10$^{34}$~erg~s$^{-1}$)  sometimes observed in persistent accreting pulsars 
and in SFXTs could be due to subsonic quasi-spherical accretion onto slowly rotating pulsars, where the accretion
of matter is mediated by a quasi-static shell above the NS magnetosphere. 
In this case, the accretion rate from the quasi-static shell depends on the ability of the plasma to enter
the magnetosphere by means of plasma cooling. These authors suggest that transitions between low-luminosity states to
bright flaring activity in SFXTs could be due to transitions between two different regimes of plasma cooling:
from thermal plasma cooling to Compton cooling dominated regime.
Unfortunately, none of these models can be confirmed to date, since both the pulsation period 
and the magnetic field in \src\ (and in most of SFXTs) are unknown.

	\section{Conclusion \label{concl}}

The new \suz\ observations we have reported here
allowed us to perform an in-depth investigation of the properties of the SFXT \src\ 
along one orbital cycle, and to obtain the following results:
\begin{itemize}
\item the source spends most of the observation with an average out-of-eclipse (intra-flares)
X--ray luminosity of 10$^{34}$~erg~s$^{-1}$, with several kinds of variabilities on different 
timescales, but with 
no bright flares (exceeding 10$^{36}$~erg~s$^{-1}$) during the observed orbital cycle. Given 
long-term monitoring performed with previous missions, this is very likely the most typical 
appearence of an orbital cycle in this SFXT;

\item the absorbing column density does not show evidence for variability, within the uncertainties,
except from during the X--ray eclipse;

\item the remaining flux during the X--ray eclipse, produced by Thomson scattering, 
allowed us to estimate the wind density at the orbital separation, 
resulting in $\rho$$_{\rm w}(a)$=7$\times$10$^{-14}$~g~cm$^{-3}$;

\item assuming a circular orbit and a spherical geometry for the supergiant wind, the 
derived wind density translates into a
ratio $\dot{M}_{w}/ v_{\infty}=7\times$10$^{-17}$~M$_{\odot}$/km. Since the accretion luminosity implied by this
ratio, assuming  terminal velocities in the range 500--3000~km~s$^{-1}$, is at least two orders of magnitude
higher than that observed, we can conclude that a mechanism 
mediating the accretion onto the putative neutron star
in the system is likely to be at work to reduce the mass accretion rate.

\end{itemize}

\section*{Acknowledgments}

This work is based on data from observations with \suz.
%
%
This work was supported by the grant from PRIN-INAF 2009, ``The
transient X--ray sky: new classes of X--ray binaries containing neutron stars''
(PI: L. Sidoli).
AB received funding from NASA grant 11-ADAP11-0227.
LS is grateful to Angela Bazzano and 
Tim Oosterbroek for very helpful comments and suggestions.
PE thanks Sara Motta for fruitful discussions.
This research has made use of the IGR Sources page maintained by 
J.~Rodriguez \& A.~Bodaghee (http://irfu.cea.fr/Sap/IGR-Sources/).

\bibliographystyle{mn2e} 




\bsp

\label{lastpage}

\end{document}